# The electronic structure of zircon-type orthovanadates: Effects of high-pressure and cation substitution

V. Panchal[1], D. Errandonea[1,*], A. Segura[1], P. Rodríguez-Hernandez[2], A. Muñoz[2], S. Lopez-Moreno[2], and M. Bettinelli[3]

[1]MALTA Consolider Team, Departamento de Física Aplicada-ICMUV, Universidad de Valencia, Edificio de Investigación, C/Dr. Moliner 50, E-46100 Burjassot (Valencia), Spain

[2]MALTA Consolider Team, Departamento de Física Fundamental II, and Instituto de Materiales y Nanotecnología, Universidad de La Laguna, E-38205 Tenerife, Spain

[3]Laboratory of Solid State Chemistry, DB and INSTM, Università di Verona, Strada Le Grazie 15, I-37134, Verona, Italy

**Abstract:** The electronic structure of four ternary-metal oxides containing isolated vanadate ions is studied. Zircon-type $YVO_4$, $YbVO_4$, $LuVO_4$, and $NdVO_4$ are investigated by high-pressure optical-absorption measurements up to 20 GPa. First-principles calculations based on density-functional theory were also performed to analyze the electronic band structure as a function of pressure. The electronic structure near the Fermi level originates largely from molecular orbitals of the vanadate ion, but cation substitution influence these electronic states. The studied ortovanadates, with the exception of $NdVO_4$, undergo a zircon-scheelite structural phase transition that causes a collapse of the band-gap energy. The pressure coefficient $dE_g/dP$ show positive values for the zircon phase and negative values for the scheelite phase. $NdVO_4$ undergoes a zircon-monazite-scheelite structural sequence with two associated band-gap collapses.

* Corresponding author; email: daniel.errandonea@uv.es

## I. Introduction

During the last decade, intensive investigations have been carried out on the structural evolution of the zircon- and scheelite-type $ABO_4$ compounds under extreme conditions. These compounds show diverse applications in various fields such as host materials for high-power lasers [1, 2], scintillators for γ-ray detection [3, 4], thermophosphorus sensors [5], and nuclear-waste storage medium [6]. Due to their exceptional optical properties, like wide optical transparency and large birefringence, the rare-earth orthvanadates are potential candidates for optical isolators, circulators beam displacers and components for polarizing optics. Other than technological importance these compounds have generated considerable theoretical interest due to the presence of 4f electrons. All these compounds exhibit similar zircon structure as the 4f sub-shell changes monotonically with few exceptions like $LaVO_4$. Hence from the theoretical point of view it is important to understand the possible changes in the band structure and variation in the covalent effects due to 4f electrons.

The rare-earth orthovanadates $AVO_4$ (where A = Nd, Yb, Lu, etc.) and $YVO_4$ crystallize in tetragonal zircon-type structure (space group: $I4_1/amd$, Z = 4) at ambient conditions [7, 8] as shown in **Figure 1**. In this structure the vanadium atom is tetrahedrally coordinated while the trivalent A cation is coordinated by eight oxygen atoms forming a bidisphenoid. Recently, several high-pressure investigations like x-ray diffraction [9-12], Raman scattering measurements [13-17], and theoretical calculations [12, 17, 18, 19] have been carried out to understand the structural modifications induced by pressure. According to these investigations the general trend for the family of these compounds suggest that the zircon-type vanadates transform generally to a much denser scheelite phase (space group: $I4_1/a$, Z = 4, as shown in **Fig. 1**) with a volume collapse of ~10 % due to more efficient polyhedral packing [9, 20]. The denser scheelite phase further transform to monoclinic fergusonite phase (space group: I2/a) [16]. The high-pressure structural phase transitions are well documented by many of the investigations mentioned earlier and the mechanism of structural phase transition is also well understood.

However, very few high-pressure optical-absorption studies have been carried out for the zircon and scheelite compounds [14, 21]. In fact, except few zircon [22] and scheelite [23] oxides there is no knowledge of the optical band-gap energy ($E_g$) at ambient conditions for most of them. As these compounds show exceptional optical properties the study of band-gap energy and its behavior at extreme conditions becomes crucial from the application point of view. In addition, as mentioned earlier these compounds involve 4f sub-shell electrons and it is interesting to know the role played by them on the electronic structure of orthovanadates and the effect of high pressure on the evolution of the electronic structure. The motivation of the present work is to understand the effect of high pressure on the optical band-gap for the series of orthovanadates and to develop a theoretical understanding pertaining to the electronic structure under high pressure. Thus in the present investigations we have carried out high-pressure optical-absorption measurements in the ultraviolet-visible-near infrared region for $YVO_4$, $YbVO_4$, $LuVO_4$, and $NdVO_4$ up to 20 GPa. We have also performed reflectance measurements at ambient conditions to accurately determine $E_g$. Finally, in order to interpret the experimental results, we carried-out first-principle total-energy calculations and band-structure calculations. We used the density-functional theory (DFT) within the generalized-gradient approximation (GGA). Calculations were also performed for $BiVO_4$ and $ScVO_4$ to compare with the other compounds and facilitate a much deeper understanding of the electronic properties of the whole family of vanadates.

## II. Experimental Details

The $AVO_4$ samples (A = Y, Lu, Nd, Yb) used in the experiments were obtained from single crystals prepared by the flux growth method using $Pb_2V_2O_7$ as the solvent [24]. Appropriate quantities of pure $V_2O_5$, PbO, $Na_2B_4O_7$, and 99.99% $A_2O_3$ were used as starting materials. $Na_2B_4O_7$ was added as flux modifier to increase the size of the crystals. After careful mixing the starting mixtures were put in Pt crucibles and heated to 1270 °C in a horizontal programmable furnace. The melts were maintained at this temperature for 12 h (soaking time), then cooled to 800 °C at a rate of 1.8 °C $h^{-1}$. The crucibles were then drawn out from furnace and quickly inverted to separate the flux from the crystals grown at the crucible bottom. Transparent crystals having an average size 3 × 2 ×1 $mm^3$ and elongated in the direction of the crystallographic *c*-axis of the tetragonal zircon structure were separated from the flux by dissolving it in hot diluted $HNO_3$. The zircon structure of each crystal was confirmed by x-ray diffraction measurements.

For the optical-absorption measurements, small single crystals of size ~ 80 μm × 80 μm and typical thickness of 10 - 20 μm were cleaved along the {110} plane. These measurements were performed at ambient pressure and upon compression. For the high-pressure studies the crystals were loaded in a 180 μm hole of an Inconel gasket pre-indented to 50 μm in a membrane diamond-anvil cell (DAC). The culet-size of the IIA-type diamond anvil was 480 μm. Small ruby balls were loaded together with sample for pressure determination [25]. Methanol-ethanol-water (16: 3: 1) mixture was used as pressure-transmitting medium. For the reflectance measurements we have used directly a polished surface of the large single crystals. The high-pressure optical-absorption measurements were carried out in the ultraviolet (UV)-visible (VIS)–near-infrared (NIR) range using an optical set-up consisting of a deuterium lamp, fused silica lenses, reflecting optics objectives, and an UV-VIS-NIR spectrometer. This set-up allows transmission measurements up to 5.5 eV [26]; i.e. up to higher energies than the absorption edge of IIA diamonds. The optical-absorption spectra were obtained from the transmittance spectra of the sample which were measured using the sample-in sample-out method [27]. The reflectance measurements were carried out at normal incidence.

**III. Overview of the Calculations**

Total-energy calculations were performed with the density-functional theory (DFT) [28], the plane-wave method, and the pseudopotential theory with the Viena ab initio simulation package (VASP) [29]. We use the projector-augmented wave scheme (PAW) [30] implemented in this package. Basis set including plane waves up to a energy cutoff of 520 eV were used in order to achieve highly converged results and accurate description of the electronic properties. Dense special k-points sampling for the Brillouin zone integration were performed in order to obtain very well converged energies and forces, within 1-2 meV/atom for the energy and smaller than 0.006 ev/Å for the forces. We used the generalized-gradient approximation (GGA) for the description of the exchange-correlation energy with the Perdew-Burke-Ernzerhof (PBE) [31] prescription (Also the local-density approximation, LDA, was used in some compounds in order to check that our results are qualitatively similar for different functionals). The application of DFT-based total-energy calculations to the study of semiconductors properties under high pressure has been reviewed [32], showing that the phase stability, electronic and dynamical properties of compounds under pressure are well described by DFT. Also this method has been applied to the study of several $ABO_4$ compounds [33, 34]. We exclude of our study the theoretical analysis of $YbVO_4$ because there is not available a good pseudopotential to describe the f electrons of the Yb atom. It is well known that DFT could yield incorrect results for compounds with small f orbital overlap and narrow f bands. We found that this is the case for $YbVO_4$ (in opposition to $LuVO_4$ and $NdVO_4$), where calculations do not reproduce accurately the ambient pressure crystalline structure of it. Therefore, we will focus our theoretical study on $YVO_4$, $NdVO_4$, and $LuVO_4$. Calculations were also performed for $BiVO_4$ and $ScVO_4$ to provide a broader picture of the electronic structure of vanadates.

## IV. Experimental results

In order to determine the ambient condition band-gap, $E_g$, for the orthovanadates, we have employed light reflectance and absorption measurements in the UV-VIS-NIR region. The corresponding reflectance spectra for all these vanadates are shown in **Fig. 2**. The reflectance spectra show a broad hump with a maximum around ~ 3.8 eV and a sharp maximum close to ~ 4.4 eV. As can be seen in the inset of **Fig. 2** for $YVO_4$, the first maximum of the reflectivity is found to overlap with the absorption edge indicating that it corresponds to the onset of allowed step-like direct transitions and can reasonably be assigned to the direct band-gap. By fitting the reflectance maximum with a Gaussian fit we have calculated the band-gap energy for all the vanadates and it has been tabulated in **Table I**. All reflectivity spectra exhibit another maximum at 4.4 eV followed by a minimum at 4.6 eV (slightly shifted to lower energies in $NdVO_4$). This maximum-minimum structure in reflectivity reveals the presence of an intense absorption peak. We can reasonably scale the arbitrary reflectivity units of **Fig. 2** by assuming that the reflectivity value (R) around 3 eV corresponds to a refractive index of 2, what would yield an absolute value R = 0.11. With this absolute scale, the reflectivity peak at 4.4 would correspond to a refractive index of 2.8 which, in turn would imply an extinction index of the same order corresponding to an absorption coefficient larger than $10^6$ $cm^{-1}$. The high intensity of this peak would reveal a peak of a joint density of states.

In **Fig. 3** we show the absorption spectra of $YVO_4$ and $LuVO_4$ at selected pressures. Since all the compounds show a qualitative similar behavior, for the sake of briefness, we only represented the results for these two compounds. The typical values of the absorption coefficient were of the order of 2000-2500 $cm^{-1}$. These values are characteristic for the low-energy tails of direct-absorption edges. The four studied compounds exhibit a steep absorption characteristic of a direct band-gap. This absorption edge exhibits an exponential dependence on the photon energy following Urbach's law [35]. Therefore, in order to follow $E_g$ under compression, we analysed the absorption spectra assuming that a band-gap of direct type (our calculations also indicate that orthovanadates are direct-gap semiconductors) and that the absorption edge obeys [35]:

$$\alpha = A_0 e^{\frac{-(E_g - h\nu)}{E_U}} \qquad (1)$$

In this equation $E_U$ is the Urbach's energy, which is related to the steepness of the absorption tail, and $A_0 = k\sqrt{E_U}$ for a direct band-gap [36], where *k* is the characteristic parameter of each material. By fitting the above equation to the measured absorption spectra at ambient conditions we have obtained the values of $E_g$. We present them in the **Table I**. As it can be seen, all the compounds have band-gap energies close to ~ 3.8 eV and the values obtained are in quite good agreement with those obtained from reflectivity measurements. The $E_g$ value we obtained for $YVO_4$ agrees with previously reported values [22, 37, 38]. The band-gap energy of the four studied compounds is also similar to that reported for $LaVO_4$ from diffuse reflectance measurements [22] and those deduced for $GdVO_4$ and $LuVO_4$ from the excitation spectra [39].

Among the four studied orthovanadates $Yb^{3+}$ and $Nd^{3+}$ cations have partially filled 4f orbitals, $Y^{3+}$ has no f electrons, and $Lu^{3+}$ has a complete filled f-shell. It has been argued that the contribution of 4f electrons either to the valence or conduction band could lead to a reduction of $E_g$ [22]. This hypothesis was based upon diffuse reflectance measurements given $E_g$ = 1.8 eV for $CeVO_4$ [22]. In contrast, from optical absorption measurements $E_g$ = 3.2 eV was obtained [40]. Our results show that $YVO_4$, $LuVO_4$, $NdVO_4$, and $YbVO_4$ have a quite similar $E_g$, which compares very well with band gap of $LaVO_4$ [22, 41] and $GdVO_4$ [39]. In addition, optical-absorption measurements gave an $E_g$ slightly smaller for $CeVO_4$ which is very similar to the band gap of $ScVO_4$ [42]. This similitude suggests that 4f electrons play a secondary role in the band structure of rare-earth orthovanadates. Indeed, as we will show in our calculations the V-O interaction dominates the electronic properties of these compounds.

**Fig. 4** shows the variation of the $E_g$ versus pressure up to 20 GPa for the four studied vanadates. There is a small linear increase in the band-gap energy with increasing pressure for all compounds, being the pressure coefficient approximately 18 meV/GPa. The only exception is $NdVO_4$, which exhibits a much lower value (some 9 meV/GPa). The pressure coefficients are summarized in **Table II**.

In the case of $YVO_4$ we have observed a sudden drop in $E_g$ around 7.4 GPa indicative of a structural phase transition. This experimental finding is in agreement with the earlier reported high-pressure x-ray diffraction [10] and Raman scattering [13] measurements. Indeed the zircon-scheelite transition occurs around 7.5 GPa. The drop of 1.05 eV in $E_g$ at the transition pressure is consistent with the color change detected in earlier optical measurements [13]. For the high-pressure phase we have observed a linear decrease of $E_g$ with pressure (see **Table II** and **Fig. 4**). We did not find any other abrupt change in $E_g$ up to 20 GPa.

In the case of $YbVO_4$ we have observed similar features for the variation in the band-gap energy with pressure. Beyond the pressure of ~ 7.3 GPa we have observed a collapse of $E_g$ from 3.9 to 2.8 eV, this change of 1.1 eV is of the same order than the $E_g$ change found in $YVO_4$ (1.05 eV). This discontinuity in $E_g$ is consistent with the zircon to scheelite transition [43]. The pressure coefficients of $E_g$ for the low- and high-pressure phases of $YbVO_4$ are summarized in **Table 2**. Again, in the low-pressure phase the band gap opens upon compression and in the high-pressure phase the opposite behavior is found. On further increase of pressure till 17.4 GPa we did not observed any abrupt change in the band-gap energy $E_g$.

In the case of $LuVO_4$ initially the band-gap energy increases linearly with pressure. This increase of $E_g$ in the zircon phase is found to be 18.9 meV/GPa. However, beyond the pressure of ~ 8 GPa we have observed a collapse of $E_g$ from 3.95 to 2.94 eV. This collapse in $E_g$ is comparable to those found in $YVO_4$ and $YbVO_4$. The change of $E_g$ is again consistent with the zircon to scheelite phase transition [15]. Beyond 8 GPa the pressure evolution is similar to the scheelite phase of the other vanadates. The rate of decrease of $E_g$ in the high-pressure phase is found to be -19.7 meV/GPa. This value is similar to that of $YVO_4$ and $YbVO_4$ and also to that previously found for $TbVO_4$ and $DyVO_4$ [14]. In **Fig. 3**, the pressure dependence of the band-gap energy $E_g$ is shown till 15 GPa as beyond this pressure the experimental data were not reproducible due to development of defects in the crystal probably caused by the deterioration of quasi-hydrostatic conditions beyond 15 GPa. However, the absorption spectra have been measured till 20 GPa. We did not see any evidence of the second phase transition reported in the Raman investigation by Rao *et. al.* (16 GPa) [15]. This is consistent with Errandonea *et al.* finding the scheelite-fergusonite transition at 21 GPa [9].

For zircon $NdVO_4$, whose band-gap exhibits a low pressure coefficient, a different behavior is detected. Around the pressure of ~ 6 GPa we have seen a drop in $E_g$ from 3.72 to 3.30 eV, which is ~ 60 % smaller than the one observed in $YVO_4$, $YbVO_4$, and $LuVO_4$. This could be attributed to a structural phase transition at this pressure. Preliminar Raman spectroscopy measurements indicate that the zircon to monazite transition occurs around 6 GPa [44]. Our calculations also support this conclusion. Apparently, $NdVO_4$ behaves under compression in a distinctive way than most vanadates, exactly like $CeVO_4$ does [45]. In the pressure range from 6 to 11.4 GPa the variation in the $E_g$ is observed to be linear with a slope of -16.4 meV/GPa. Additionally, beyond 11.4 GPa we have observed another drop in $E_g$ of ~ 0.5 eV. This fact indicates the presence of another phase transition in $NdVO_4$ at this pressure. According to our calculations, it corresponds to a monazite-to-scheelite transition. Note that both gap collapses together make a total-energy change close to the one observed at the zircon-scheelite transformation in the other compounds, which makes reasonable the structural sequence here proposed for $NdVO_4$. The evolution of $E_g$ beyond 11.4 GPa is observed to be linear and the rate of $dE_g/dP$ is found to be -22.2 meV/GPa. (From the **Fig. 4** it is clearly discernible the three distinct phases of $NdVO_4$.)

**IV. Theoretical results and discussion**

Total-energy calculations accurately describe the structure of different vanadates at ambient pressure. They also found the same structural sequence reported by experiments under compression. In this paper, we optimized the different structures of the studied compounds at different pressures and use the obtained parameters to calculate electronic band structures.

## A. Zircon structure

Band dispersions for zircon $YVO_4$ are plotted, along symmetry directions within the Brillouin zone (**see Fig. 5**), in **Fig. 6(a)**. The partial and total densities of states are shown in **Fig. 7**. The shapes of the bands for zircon-type $YVO_4$, $LuVO_4$, and $NdVO_4$ are very similar to each other. Therefore, we will discuss here the band structure of zircon-type $YVO_4$ and latter comment on the slight differences between this compound and the rest. According to our calculations in $YVO_4$, the valence-band maxima and conduction-band minima are located at the Γ point, so that zircon $YVO_4$ is a direct-gap material. The calculated band gap is 2.85 eV. The underestimation of the experimental $E_g$ is comparable with the typical differences between density-functional theory and experiments ($E_g$ experimental = 3.8 eV). Indeed, similar differences are found for $E_g$ in $BiVO_4$ [22] and $LaVO_4$ [22]; see **Table I**. According to calculations $dE_g/dP$ = 6 meV/GPa. This pressure coefficient is positive, like the one obtained from experiments, but its value is much smaller than the experimental one (18.7 meV/GPa). A possible origin of this discrepancy is the fact that in calculations the bulk modulus is overestimated [16], being therefore the reduction of the V-O bonds under compression underestimated. Regarding other features of the band structure, we found that the dispersion of the valence bands is relatively small, with comparable dispersions along both the *a*- and *c*-axis. We also determined that the lower half of the valence band is basically composed by V 3d – O 2p bonding states and that the upper half of the valence band consists of purely non-bonding O 2p states. On the other hand, the conduction band has antibonding V 3d – O 2p character. The conduction bands are quite narrow reflecting the large separation between $VO_4^{3-}$ units and the poor overlap of yttrium orbitals with the antibonding states of the vanadate units. The yttrium 4d states are empty and do not contribute significantly near the Fermi level. Similar conclusions have been reported recently for $ScVO_4$ [42], a compound isomorphic to $YVO_4$. Regarding the effect of pressure on the band-structure of zircon $YVO_4$, we concluded that the small increase of $E_g$ with pressure is a consequence of the small reduction of the V-O distances ($VO_4^{3-}$ tetrahedra behave as nearly uncompressible units). We performed also calculations for $ScVO_4$ confirming that this material behaves in a similar way than $YVO_4$ under compression.

Calculations of the band structure of zircon-type $NdVO_4$ and $LuVO_4$ give similar results than for $YVO_4$. To illustrate this we depict the band dispersion of $LuVO_4$ in **Fig. 8(a)**. The partial and total densities of states are shown in **Fig. 7**. Again in both compounds V 3d and O 2p states dominate the upper part of the valence band and the lower part of the conduction band. The main difference is that in $NdVO_4$ and $LuVO_4$ the band gap is slightly smaller (about 0.2 eV) because 6s electrons from the rare earth slightly contribute to the conduction band reducing the band gap. In particular, the Lu and Nd 6s states hybridize with antibonding conduction band states. Apparently we did not observe the small gap reduction in the experiments. On the other hand, the fact that rare-earth orthovanadates have a similar band structure to $YVO_4$ and $ScVO_4$ confirms that basically the $VO_4$ tetrahedron is determinant in the electronic absorption process in these materials. This conclusion is reinforced by the fact zircon-type $LaVO_4$ has also a similar $E_g$. These facts also suggest that the 3.2 eV band gap reported for $CeVO_4$ [40] is probably more accurate than the 1.8 eV value more recently published [22].

Let us now comment on the smaller band gap shown by compounds like zircon-type $BiVO_4$. In this case the factor contributing to the band-gap reduction is the hybridization of Bi 6s and 6p orbitals with V 3d – O 2p states [36]. The same differences observed between $BiVO_4$ and other vanadates are found in related ternary oxides like the tungstates. In particular, $CaWO_4$, $SrWO_4$, and $BaWO_4$ have a larger band gap than $PbWO_4$ because 6s states of Pb contribute to the conduction band something that not happen in the other compounds [21]. This distinctive feature makes also $E_g$ to be more sensitive to pressure in $PbWO_4$ than in other tungstates. By analogy, we expect the band-gap energy to be more affected by pressure in $BiVO_4$ than in the rest of the vanadates. Our calculations confirm this hypothesis (see **Table II).** From the calculations we concluded that, as in $YVO_4$, $E_g$ slightly opens upon compression in $NdVO_4$ and $LuVO_4$. In the first case, the calculated pressure coefficient agrees with the experimental value. In the second case we found the same differences than in $YVO_4$. ScVO4 also behaves in a similar way having a very small $dE_g/dP$. On the contrary in $BiVO_4$ the gap closes upon compression with a much larger pressure coefficient (-22.6 meV/GPa), as it occurs in $PbWO_4$ [21].

**B. Scheelite structure**

The scheelite structure, $I4_1/a$, is a tetragonal structure related to zircon but with lower symmetry [9]. As zircon it consists of isolated $VO_4$ terahedra and $AO_8$ dodecahedra (see **Fig.1**). Band dispersions for scheelite-type $YVO_4$ and $LuVO_4$ are plotted along the symmetry directions within the tetragonal Brillouin zone (see **Fig. 5**) in **Figs. 6(b)** and **8(b)**. As happen in the zircon structure, they are very similar to each other in the different vanadates. The valence-band maxima and conduction-band minima are located at the Γ point, so that these are direct-gap materials. In the case of $YVO_4$, at ambient pressure $E_g$ = 2.81 eV, and at the transition pressure 2.77 eV. These values are similar to the experimental $E_g$ measured from the scheelite phase. So calculations also found a collapse of the band gap at the zircon-scheelite transition, but due to the underestimation of $E_g$ in the low-pressure phase, the collapse is smaller than in experiments ($E_g$ for zircon phase at the transition pressure is 2.9 eV according to calculations and 3.9 eV according to experiments). In the case of the scheelite structure, calculations also indicate the existence of an indirect gap very close to the fundamental one. The indirect gap is from the Γ point of the valence band to the M point being its energy 2.95 eV at ambient pressure and 2.97 eV at the transition pressure. As it can be seen in **Figs. 6** and **8**, most of the features of the band structure in scheelite are qualitatively similar to the ones in the zircon structure. In scheelite, the lower part of the conduction band is composed primarily of states associated with the V 3d states and it is separated by approximately 0.5 eV from the upper part of the conduction band. All the states near the Fermi level are dominated by V 3d and O 2p orbitals.

A possible reason for the band-gap collapse can be related to the atomic rearrangement at the zircon-scheelite transition. Note that both scheelite and zircon structures consist of isolated $VO_4$ tetrahedra connected by $AO_8$ dodecahedra. They basically differ in the way these polyhedra are connected [46]. The symmetry of the large cation (e.g. Y) in scheelite is $S_4$, while in zircon is $D_{2d}$. As a consequence of it, the zircon form has a symmetric region of electron localization. The smaller electronic localization in scheelite is related to the reduction of the band gap. Also the increase of the V-O distance at the transition (around 6%) [10] helps to the reduction of the gap. Basically the Coulomb attraction of the O 2p states to the $V^{3+}$ ions is reduced causing a reduction of the splitting of O 2p and V 3d states which should lead to a reduction of $E_g$. Another possible fact leading to the reduction of the band gap is a charge transfer involving O and V ions, or O and trivalent metals [14]. These hypotheses will be explored in future studies. It is interesting to note here that fergusonite-type $BiVO_4$ (a distorted version of scheelite [47]) also shows a collapse in the band gap with respect to zircon-type $BiVO_4$ [20, 48]. This confirms that the band-gap closure is inherent to the crystallographic restructuration taking place at the transition [49, 50].

Let us discus now the pressure dependence of band-gap energy in the scheelite phase. In this phase we found small discrepancies between theory and experiments. According to experiment, in the scheelite structure the gap slightly closes upon compression but calculations predict a small gap opening (see **Table II**). One possible explanation is the existence of excitonic effects that are not taken into account in DFT calculations, but are normally strong in the absorption edge of direct semiconductors. Small subtle changes in the band structure under pressure can strongly change the exciton life and, consequently, its width. An increase of the exciton width would lead to a less steep absorption tail (as it seems to be our case, see **Fig. 3**) and compensate the band gap blue shift predicted by calculations. Measurements with thinner samples would be necessary to elucidate the origin of this small discrepancy. However, differences on pressure coefficients are of the order of a few meV/GPa. Such differences are close to the error in gap determination both in experiments and calculations. Another possibility to explain discrepancies can be differences in temperature and hydrostaticity between experiments and calculations.

To bring more light into the effects of pressure on the electronic structure of scheelite-type vanadates we will compare them with the structurally analogous tungstates. As shown in the **Table II** all the vanadates in the scheelite phase show negative pressure coefficients ranging from -6 to -22 meV/GPa. On the contrary scheelite $BaWO_4$, $SrWO_4$, and $CaWO_4$ [51] show small positive pressure coefficients. However, $PbWO_4$ is the exception with large negative pressure coefficient (-71 meV/GPa) [19]. As per the DFT calculations for the scheelite tungstates [52], bottom of the valance band is mainly occupied by W 5d states while the top of the valance band is predominantly composed of O 2p states. For cations with $ns^2$ valence configuration the contribution to valence and conduction band is negligible. On the other hand for cations with $ns^2np^0$ valence configuration like $Pb^{2+}$, a significant contribution to the valence and conduction band is expected resulting in a band-gap decrease as observed for $PbWO_4$. The negative pressure coefficient is the consequence of an increase in the hybridization of states under the application of pressure. In our case, the small negative pressure coefficients suggest that pressure induces a weak hybridization of O 2p and V 3d with s orbitals of the trivalent metal. This fact is also coherent with the fact that the scheelite structure has a smaller band gap than the zircon structure.

### C. Monazite-type $NdVO_4$

Experiments show that $NdVO_4$ has a distinctive behavior in comparison to $YVO_4$, $YbVO_4$, and $LuVO_4$. It undergoes two band-gap collapses upon compression, which are associated to consecutive phase transitions. The sequence of structures in $NdVO_4$ is zircon-monazite-scheelite. The monazite-structure and its Brillouin zone are illustrated in **Figs. 1** and **5**. Monazite is monoclinic, $P2_1/n$, and consists of $VO_4$ tetrahedra and $AO_9$ polyhedra. As in zircon, in this structure the tetrahedra types are isolated, with the only difference that in monazite they are separated by irregular $AO_9$ polyhedra instead of by $AO_8$ dodecahedra (see **Fig 1**). From our calculations we obtained that in the zircon and scheelite phases the band structure of $NdVO_4$ is quite similar to that of $LuVO_4$. Therefore, in this section, for the sake of briefness, we will present results only for the monazite phase. The band structure for monazite-type $NdVO_4$ is shown in **Fig. 10**. The density of states is shown in **Fig. 11**. According to these results, monazite $NdVO_4$ is an indirect semiconductor with the absolute maximum of the valence band at the Z point of the Brillioun zone and the absolute minimum of the conduction band at the Y point. One remarkable feature of the band structure of the monazite phase is that the dispersion of the valence and conductions bands is relatively small, with comparable dispersions along different directions. Exactly as it happen with the band structure of the zircon and scheelite phases. However, in contrast with these two structures in monazite, the region close to the Fermi level is dominated not only by O 2p and V 3d states. In this case 4f electrons of Nd are partially delocalized becoming relevant. In addition, some hybridization between orbitals is present indicating the covalent bonding character of the compound. The presence of the monazite phase as an intermediate phase explain the successive collapses of the gap, since monazite is calculated to have a band-gap energy between those of the other two structures. Each collapse of the band-gap energy is associated to a sudden contraction of the unit-cell volume. Finally we also calculated the pressure evolution of $E_g$ for the monazite phase. When comparing with experiments (see **Table II**), we found the same small discrepancies than for the scheelite phase. According to experiments $E_g$ decreases a few meV/GPa, but the opposite behavior is obtained in calculations. Reasons for it have been discussed in the previous section.

## V. Concluding Remarks

High-pressure optical-absorption measurements were performed on four orthovanadates up to 20 GPa. DFT calculations were also carried out to support the experimental findings. Most zircon-type orthovanadates undergo structural phase transition to scheelite phase as reported earlier. This transition involves a collapse of the band-gap energy $E_g$. DFT calculations also predict the collapse of the band-gap, although the values are underestimated for both phases. The closing of the band-gap at the transition can be broadly attributed to two reasons. First, due to the different site symmetry the trivalent cations have in zircon in comparison with the scheelite structure. The mixing of the electronic orbitals is forbidden in the former but allowed in the latter resulting in the collapse of the band-gap. Second, due to the increase of the V-O distance at the transition, which facilitate the reduction of the gap due to weakening of attraction of O 2p and $V^{3+}$ states as a result reduces the splitting of O 2p and V 3d states and lead to a reduction of $E_g$. In the case of $NdVO_4$ a distinctive behavior is observed, occurring two collapses of $E_g$ at 5 and 11 GPa. Both collapses are associated to structural phase transitions; zircon-monazite and monazite-scheelite. The structural changes that take places at each transition are the causes of the change in the electronic properties. The band structure of monazite $NdVO_4$ is reported, being it an indirect band-gap semiconductor.

**Acknowledgments**

We acknowledge the financial support of the Spanish MCYT (Grants MAT2010-21270-C04-01/03, CSD2007-00045, and MAT2009-14144-CO3-03). Computer time was provided by Red Española de Supercomputación (RES) and MALTA cluster. S. L-M acknowledges the support of CONACyT México through a postdoctoral fellowship.

**Figure Captions**

**Figure 1:** Crystal structure of zircon, monazite, and scheelite. The big black spheres are A atoms and white spheres are V atoms. The small black spheres are O atoms. In all phases vanadium is in tetrahedral coordination with O atoms.

**Figure 2:** Reflectance spectra versus energy for various vanadates.

**Figure 3:** Absorption spectra at representative pressures for $YVO_4$ and $LuVO_4$.

**Figure 4:** Pressure dependence of absorption edge for the vanadates. Symbols: experimental results. Lines are linear fits for the data.

**Figure 5:** Brillouin zones of zircon, monazite, and scheelite.

**Figure 6:** Band-structure of zircon (a) and scheelite (b) $YVO_4$.

**Figure 7:** (color online) Total and partial densities of states for zircon-type $YVO_4$ (a) and $LuVO_4$ (b).

**Figure 8:** Band-structure of zircon (a) and scheelite (b) $LuVO_4$.

**Figure 9:** (color online) Total and partial densities of states for scheelite-type $YVO_4$ (a) and $LuVO_4$ (b).

**Figure 10:** Band-structure of monazite-type $NdVO_4$.

**Figure 11:** (color online) Total and partial density of states for monazite-type $NdVO_4$.

**Table I:** Theoretical and experimental band-gap energy of different orthovanadates. Previous published values are also included ([a]Ref. [42], [b]Ref. [39], [c]Ref. [22], [d]Ref. [41], [e]Ref. [40], [f]Ref. [36] ).

| Compound | Eg (eV) (Reflectance) | Eg (eV) (Absorption) | Eg (eV) (Emision) | Eg (eV) Theory |
|---|---|---|---|---|
| $ScVO_4$ | | | | 2.43 - 3.2[a] |
| $YVO_4$ | 3.79 | 3.78 | | 2.79 |
| $LuVO_4$ | 3.79 | 3.76 | 3.87[b] | 2.86 |
| $LaVO_4$ | 3.5[c] | | | 2.8[c] – 3.01[d] |
| $CeVO_4$ | 1.8[c] | 3.2[e] | | |
| $NdVO_4$ | 3.72 | 3.72 | | 2.86 |
| $GdVO_4$ | | | 3.87[b] | |
| $YbVO_4$ | 3.74 | 3.79 | | |
| $BiVO_4$ | | 2.8[c] | | 2.2[f] – 2.76 |

**Table II:** Pressure coefficient of $E_g$ in different phases of the studied compounds.

| | Zircon | | Monazite | | Scheelite | |
|---|---|---|---|---|---|---|
| Compound | Experiment dEg/dP (meV/GPa) | Theory dEg/dP (meV/GPa) | Experiment dEg/dP (meV/GPa) | Theory dEg/dP (meV/GPa) | Experiment dEg/dP (meV/GPa) | Theory dEg/dP (meV/GPa) |
| $YbVO_4$ | 17.6 | | | | -17.1 | |
| $YVO_4$ | 18.7 | 6.2 | | | -6.3 | 3.8 |
| $LuVO_4$ | 18.9 | 6.9 | | | -19.7 | 2.1 |
| $NdVO_4$ | 9.1 | 7.5 | -16.4 | 6.3 | -22.2 | 4.2 |
| $ScVO_4$ | | 0.7 | | | | |
| $BiVO_4$ | | -22.6 | | | | |

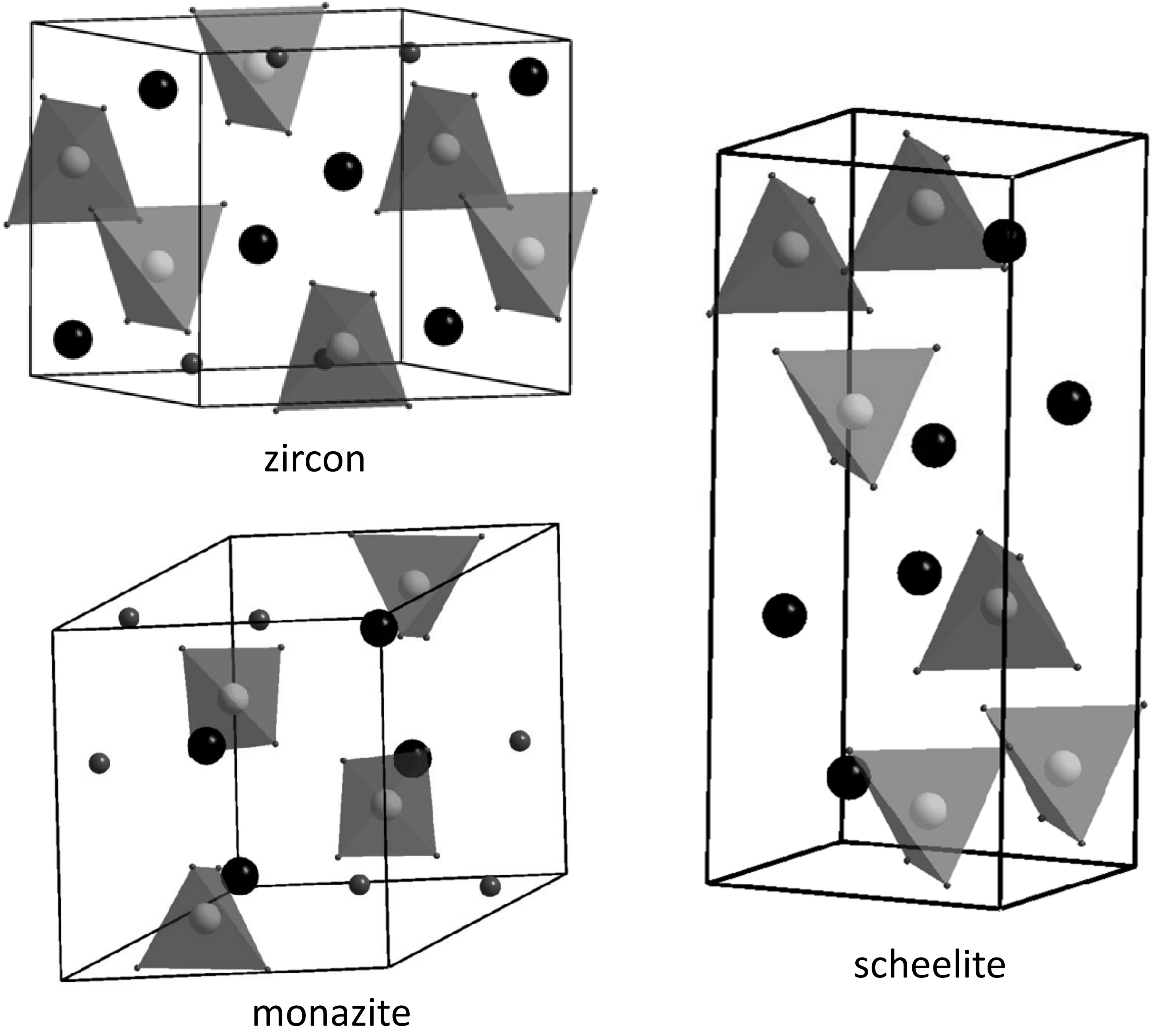
zircon
monazite
scheelite

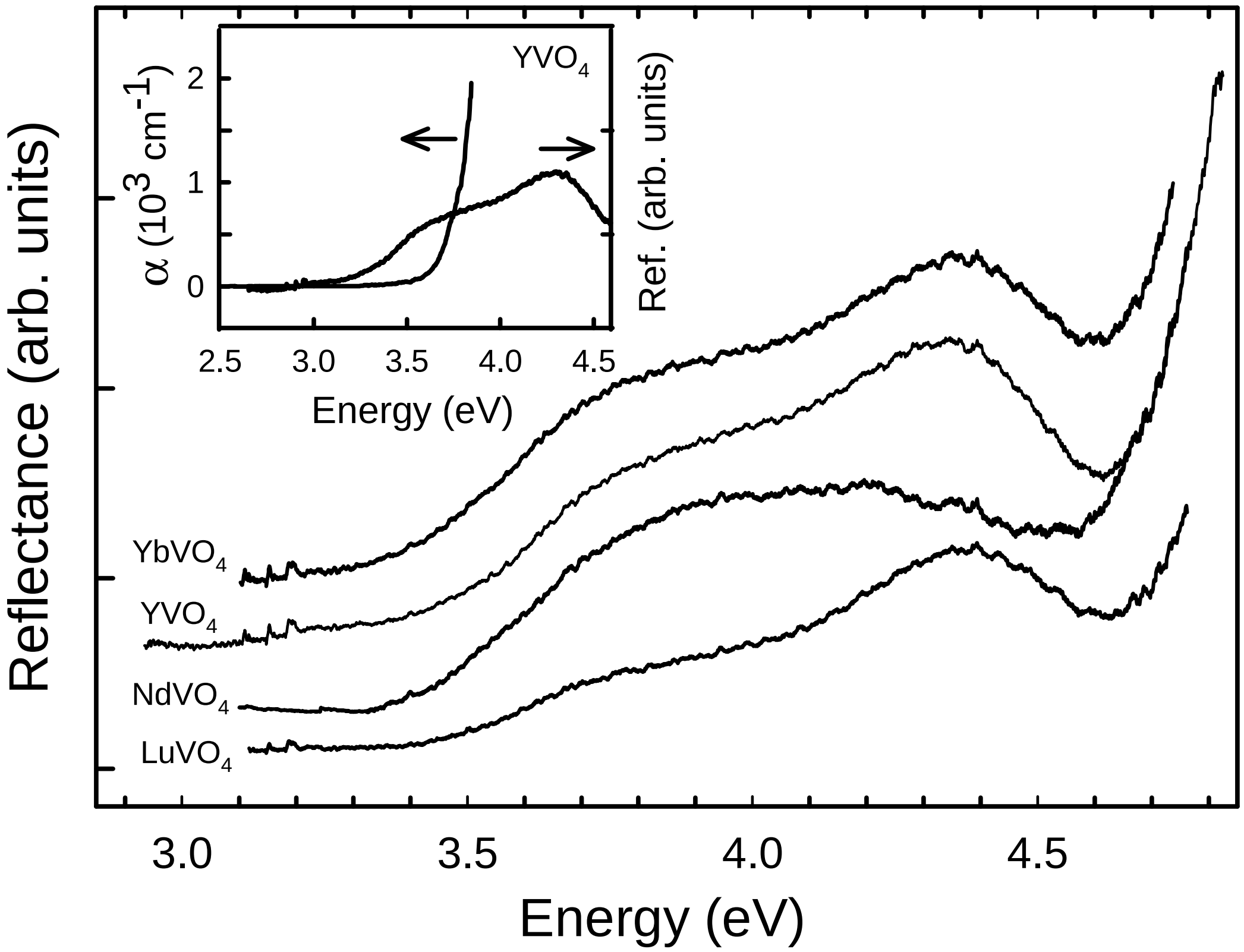

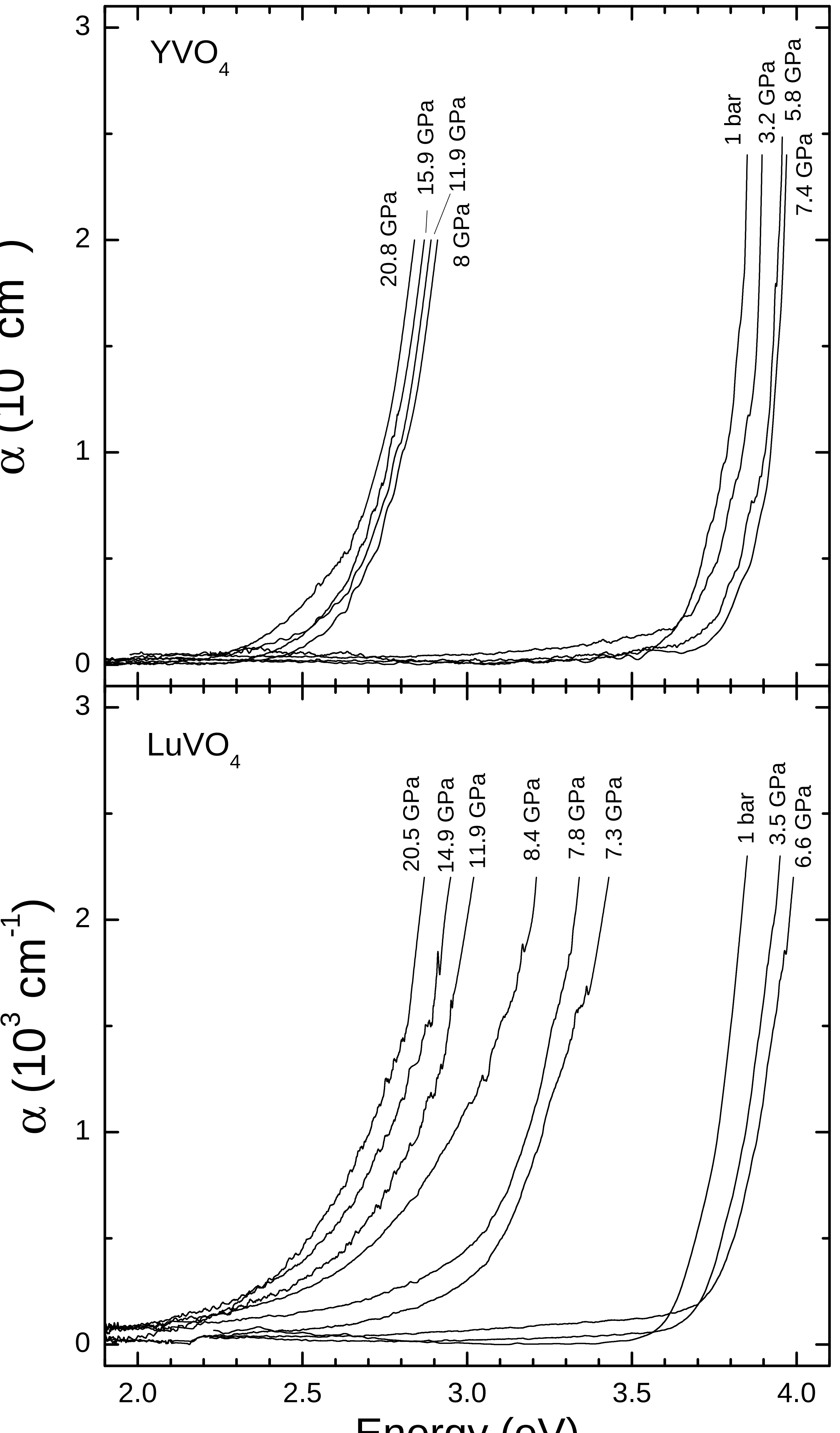

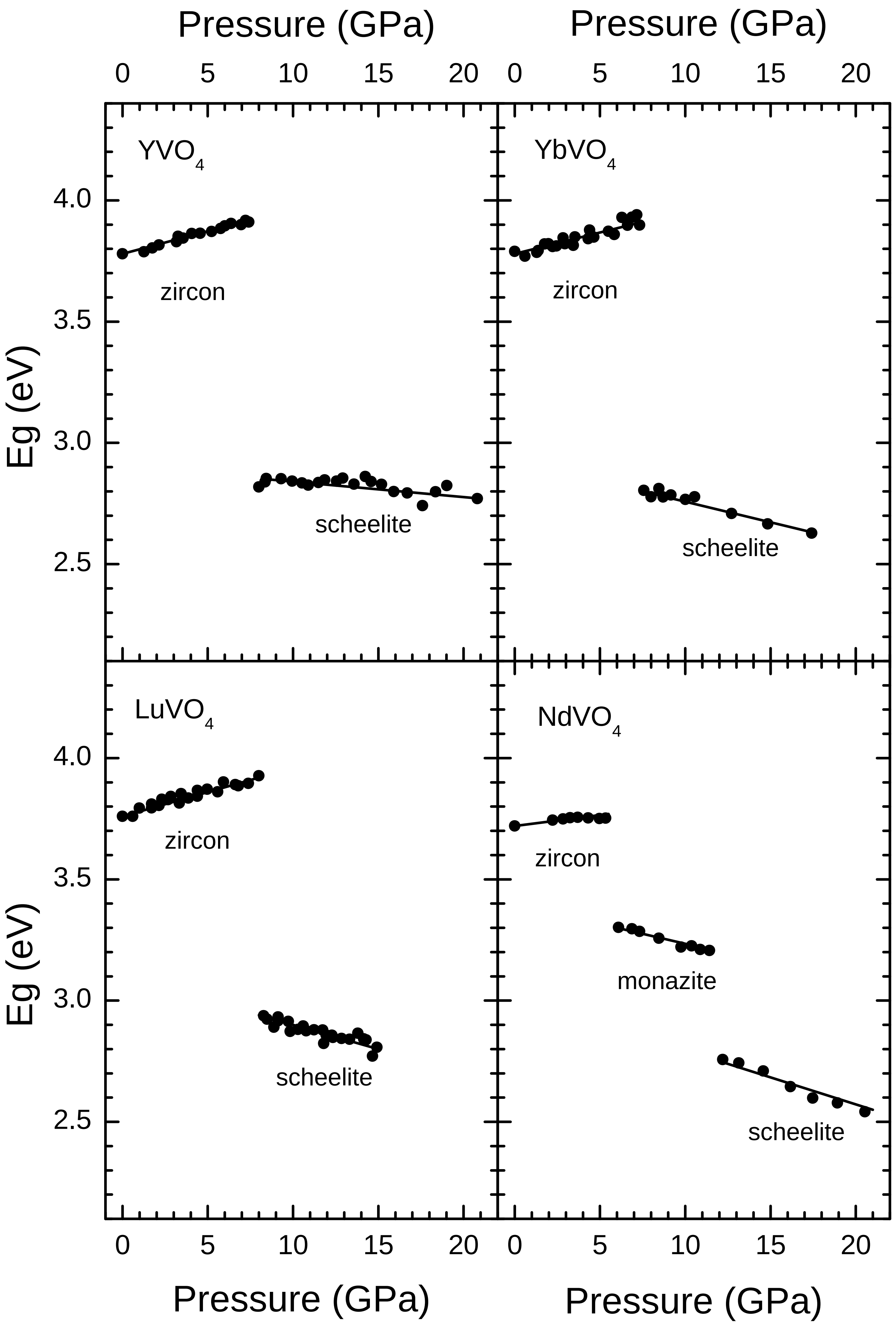

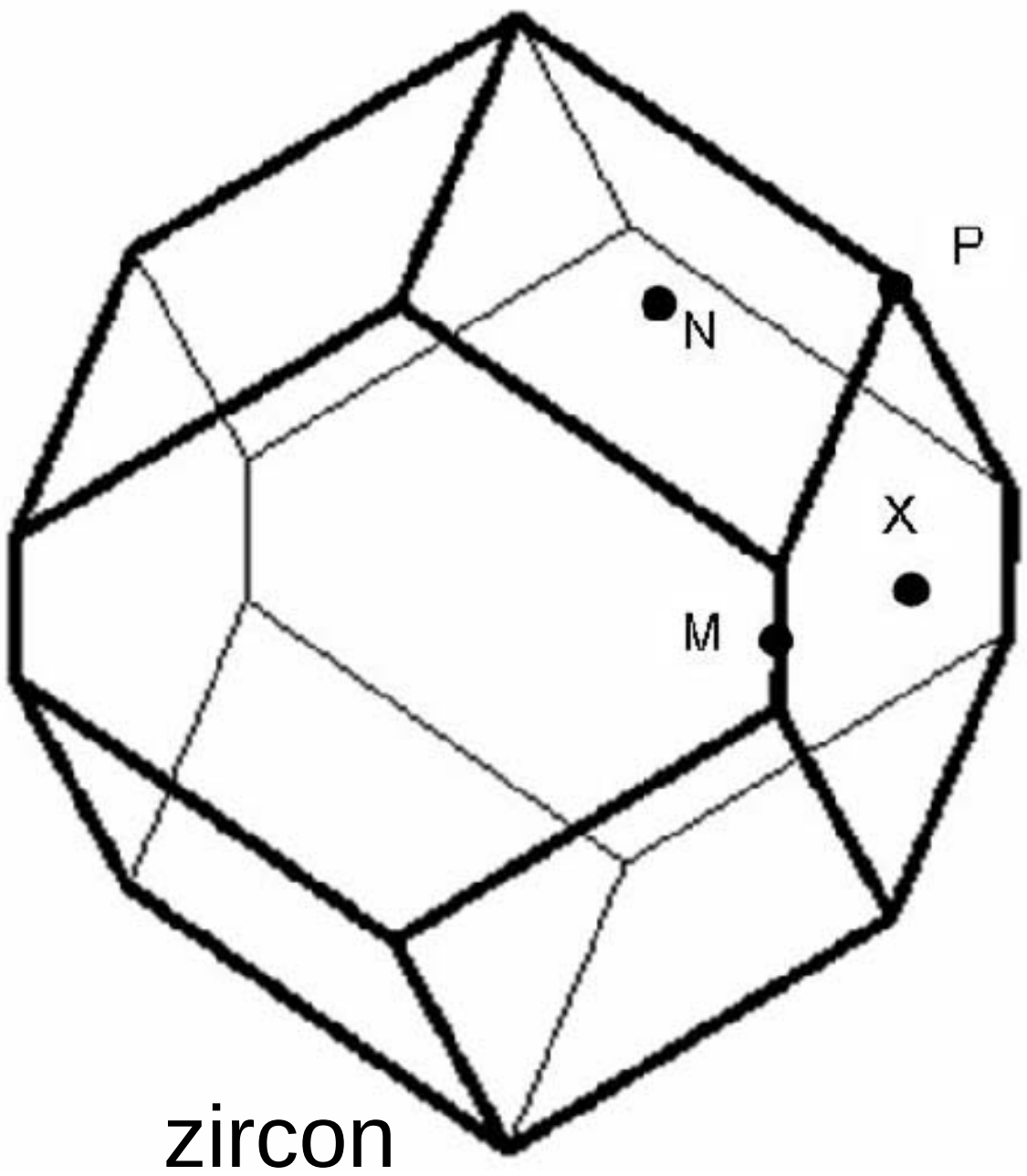

zircon

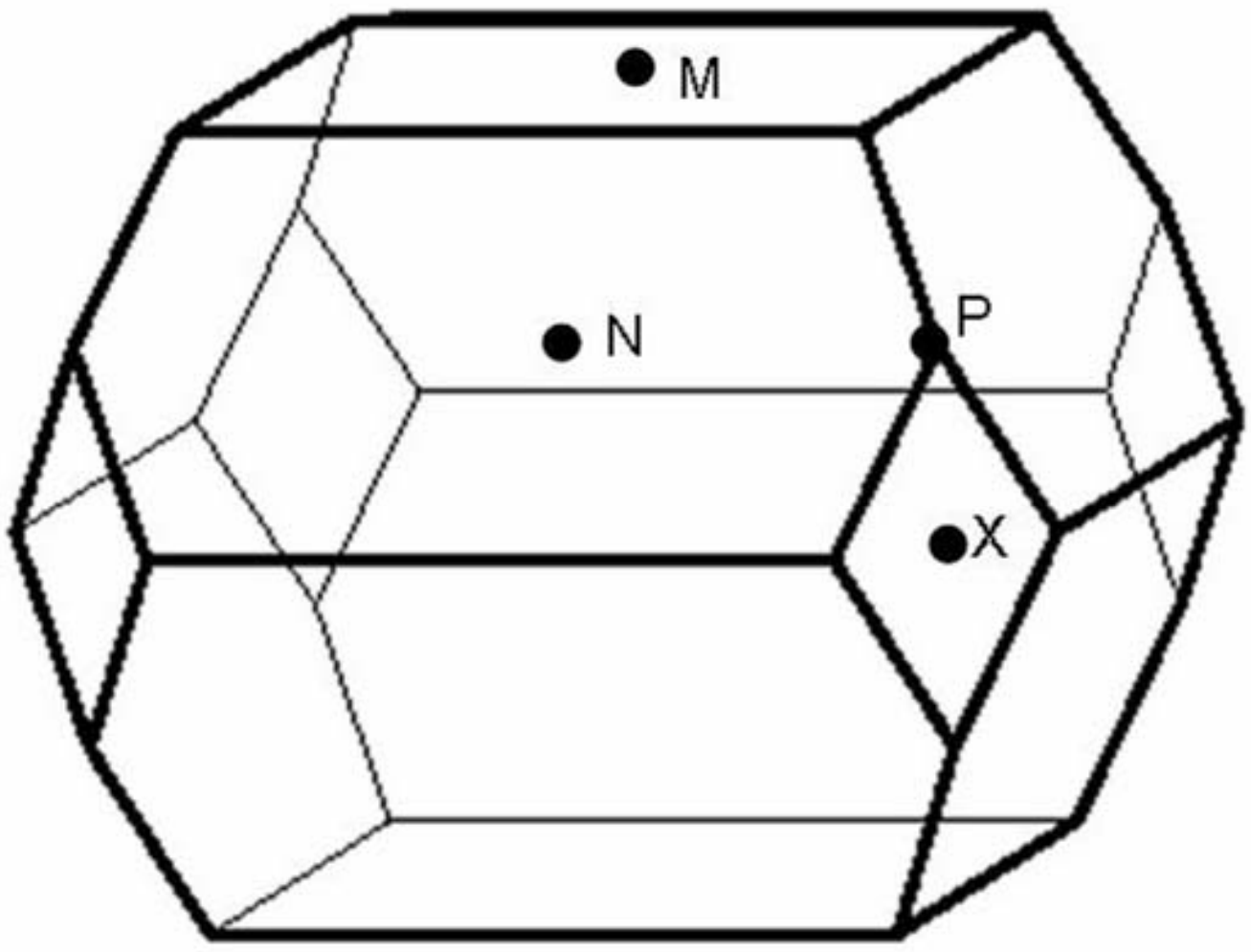

scheelite

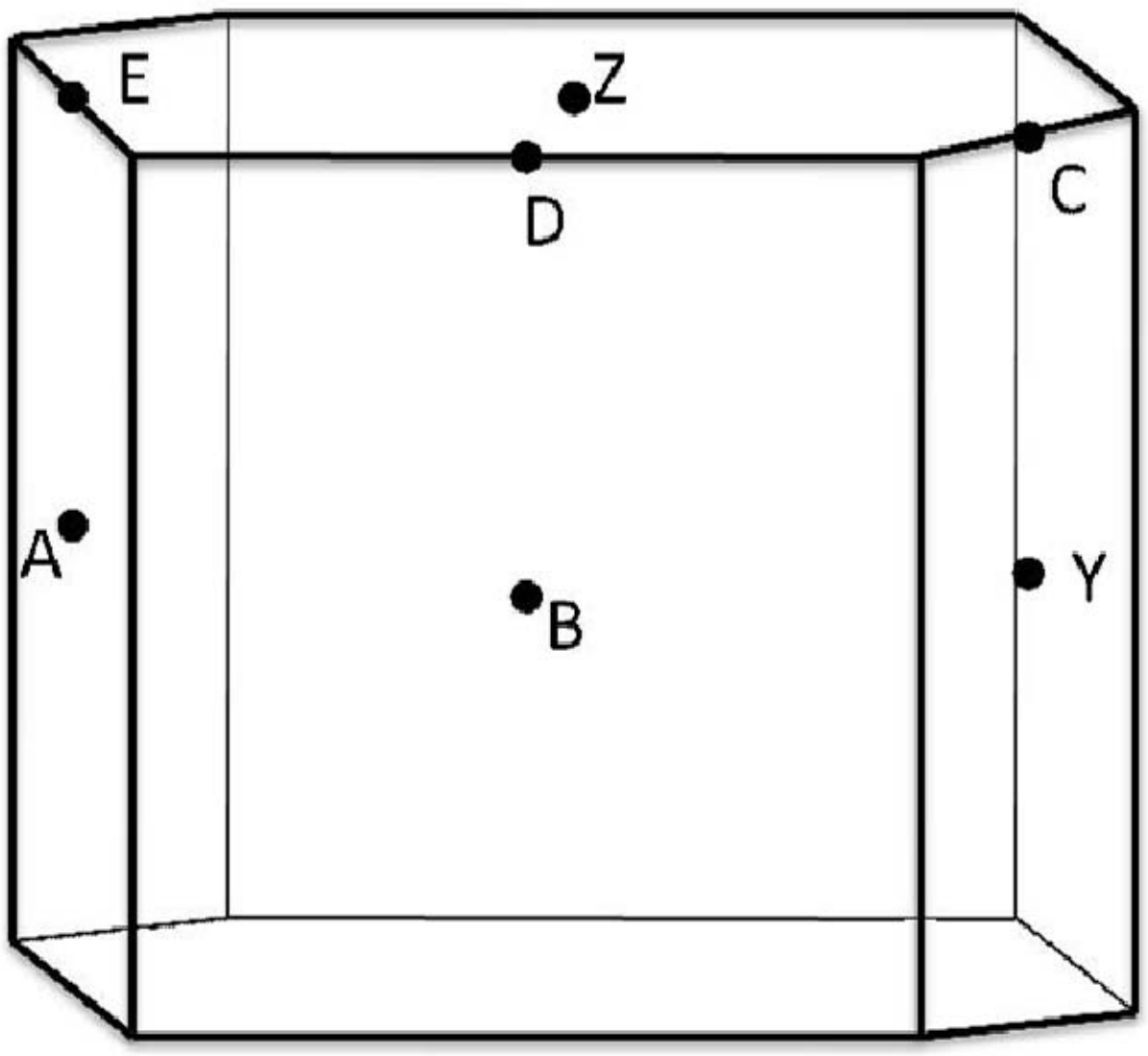

monazite

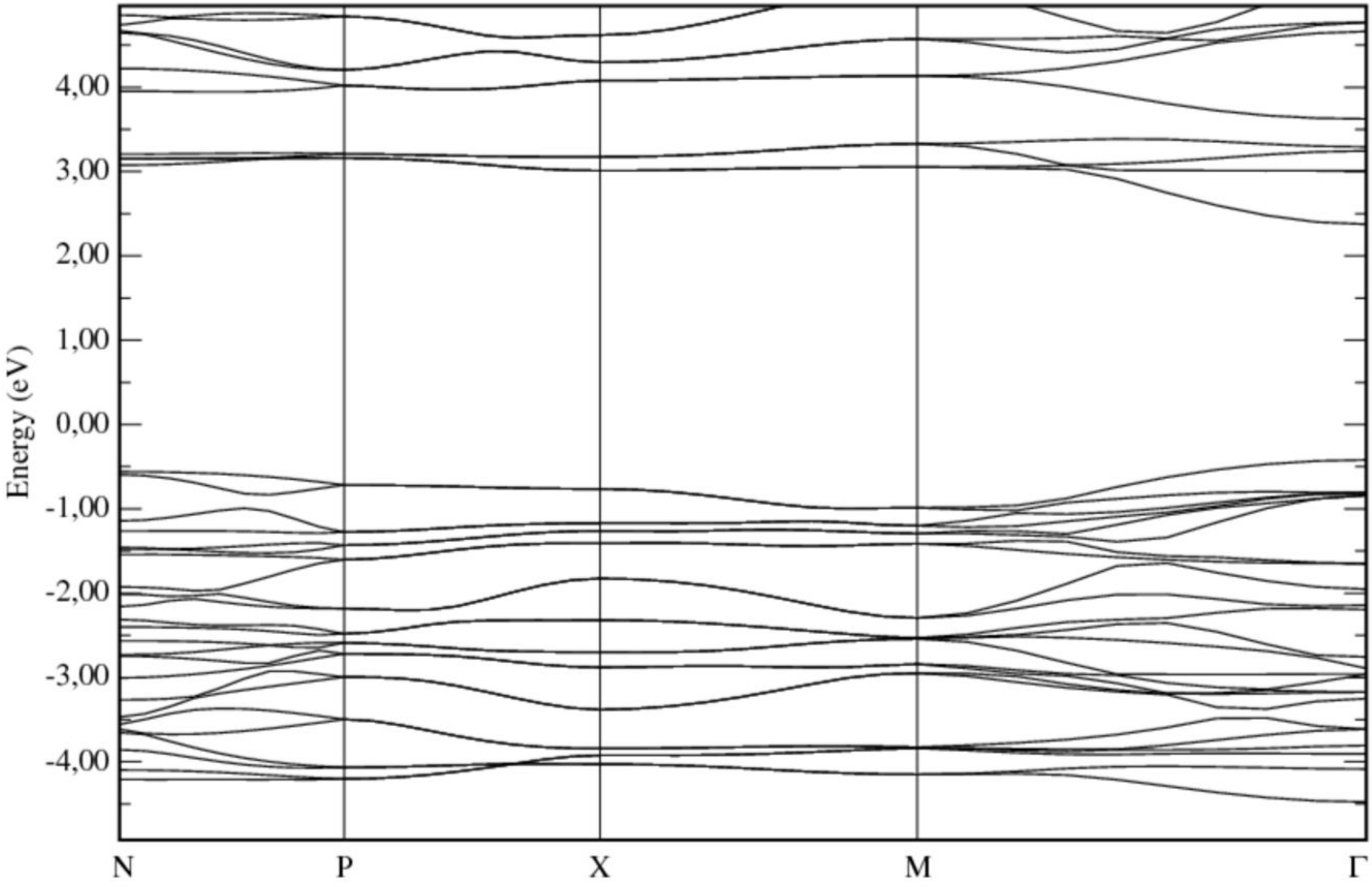

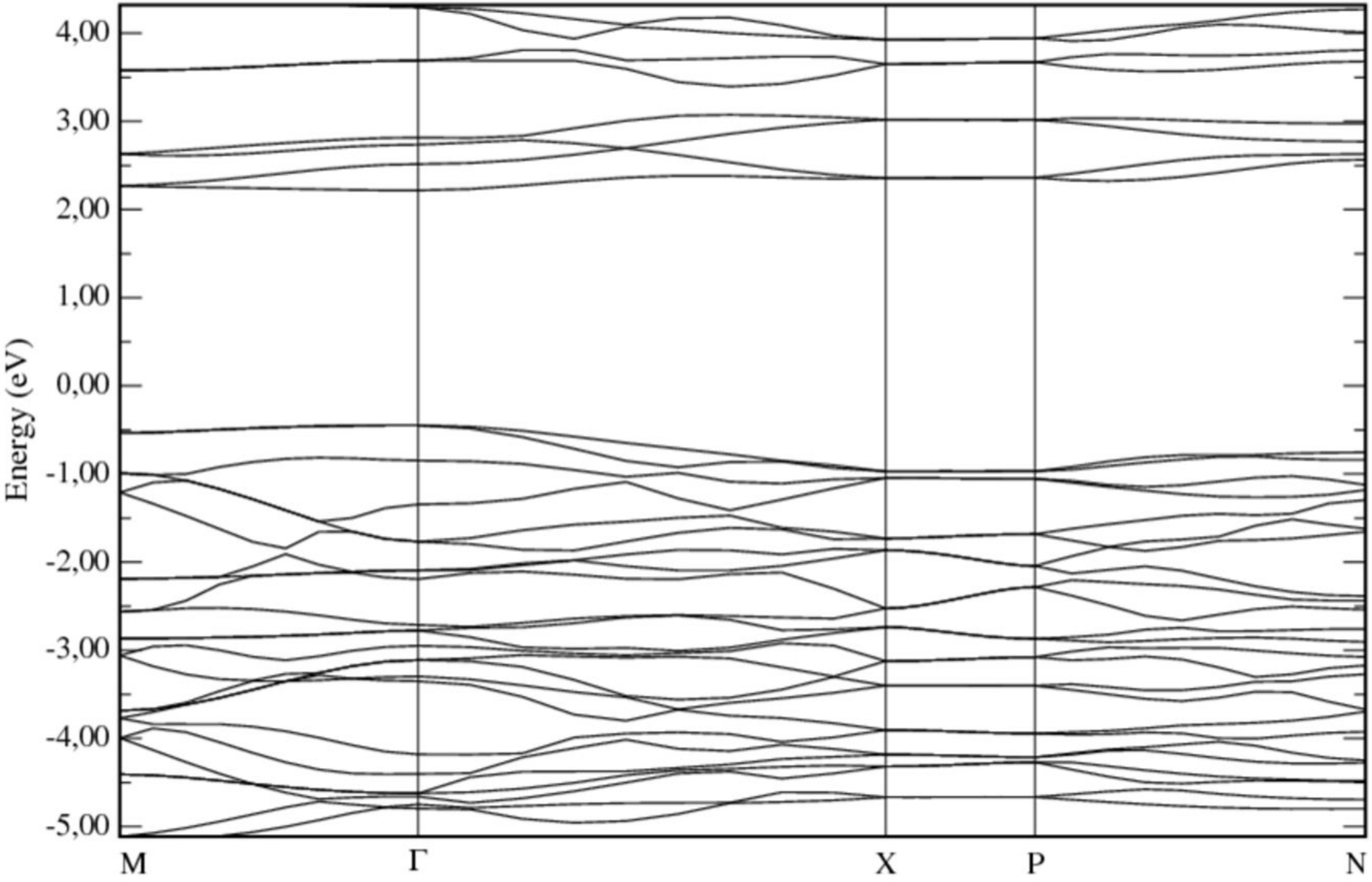

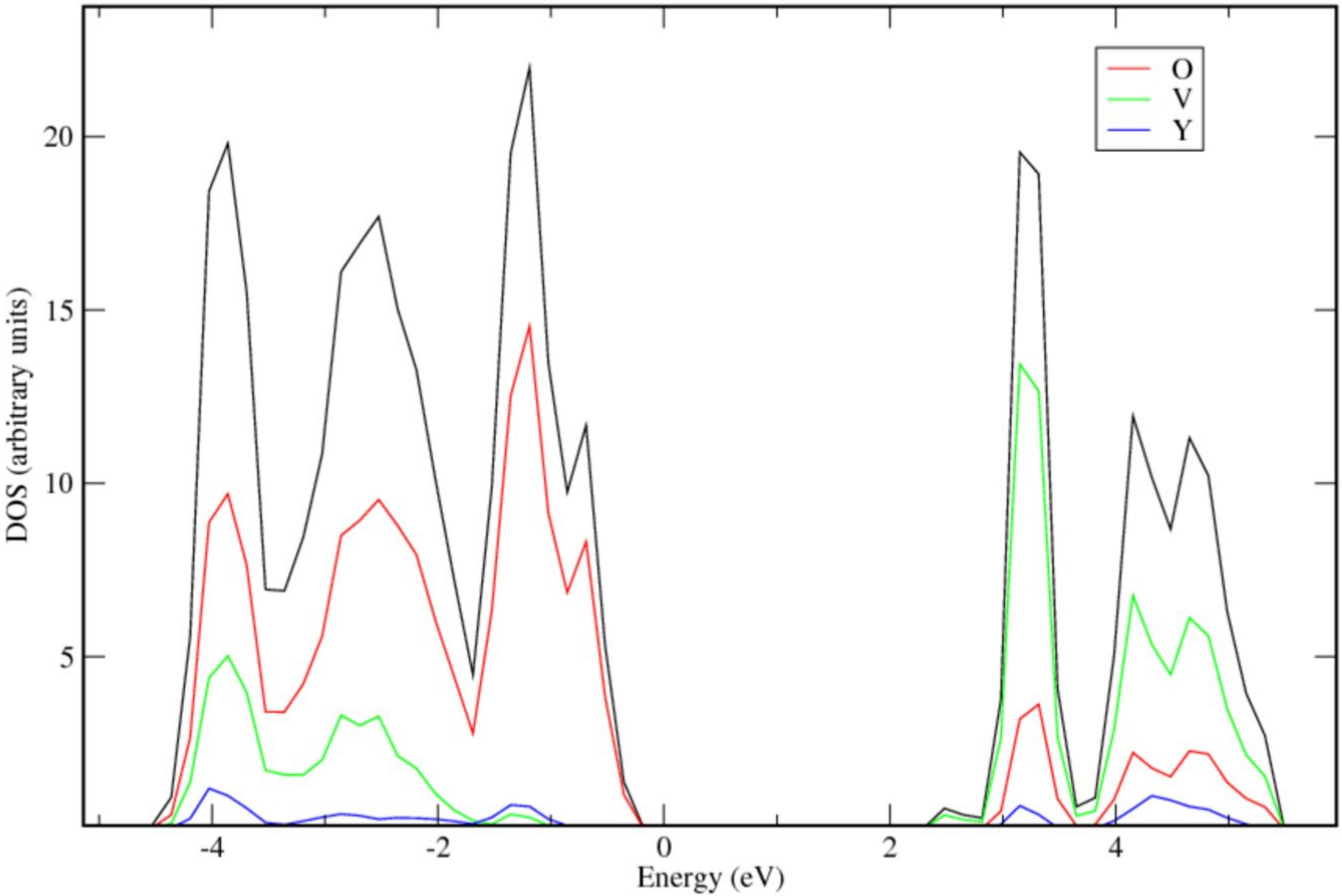

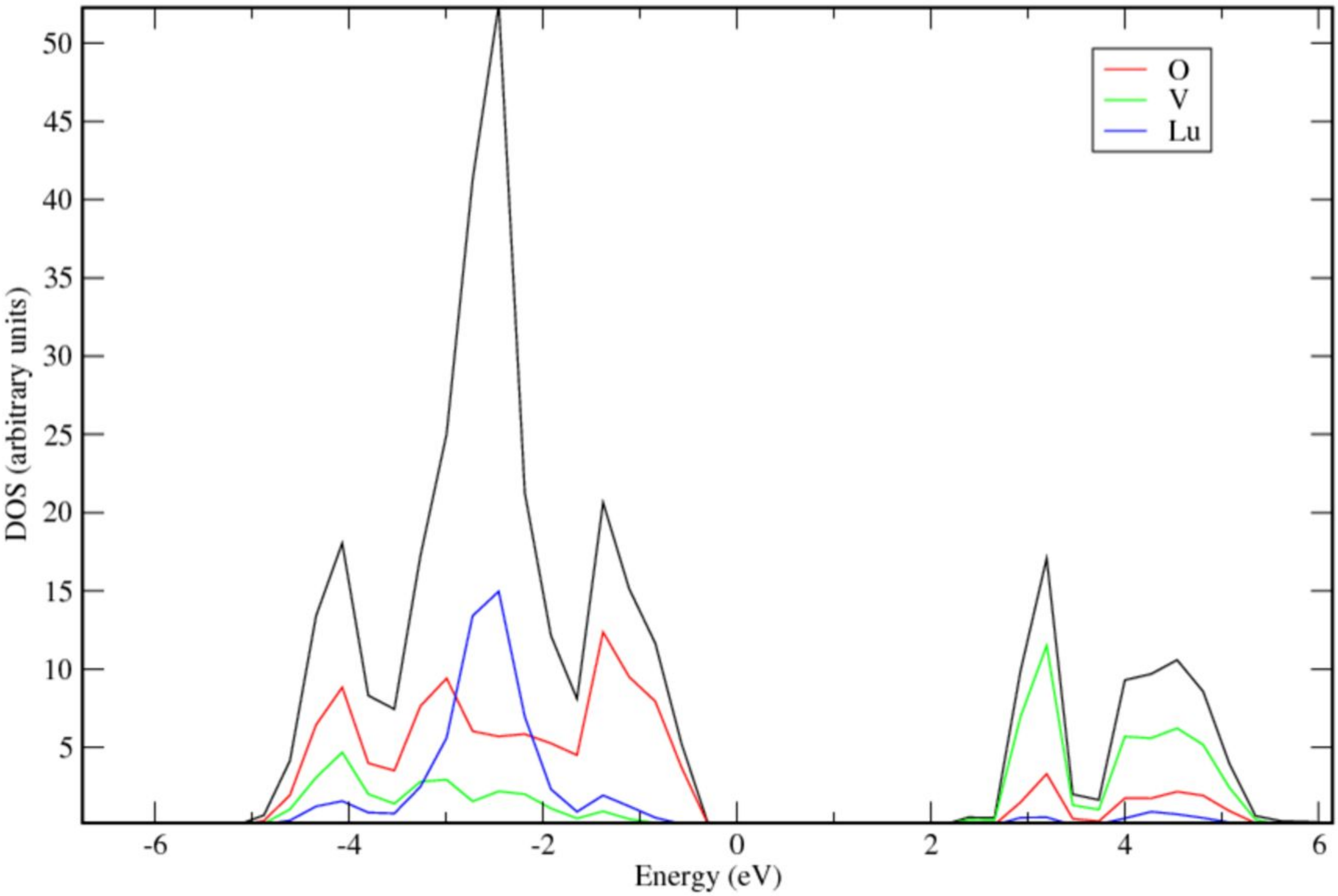

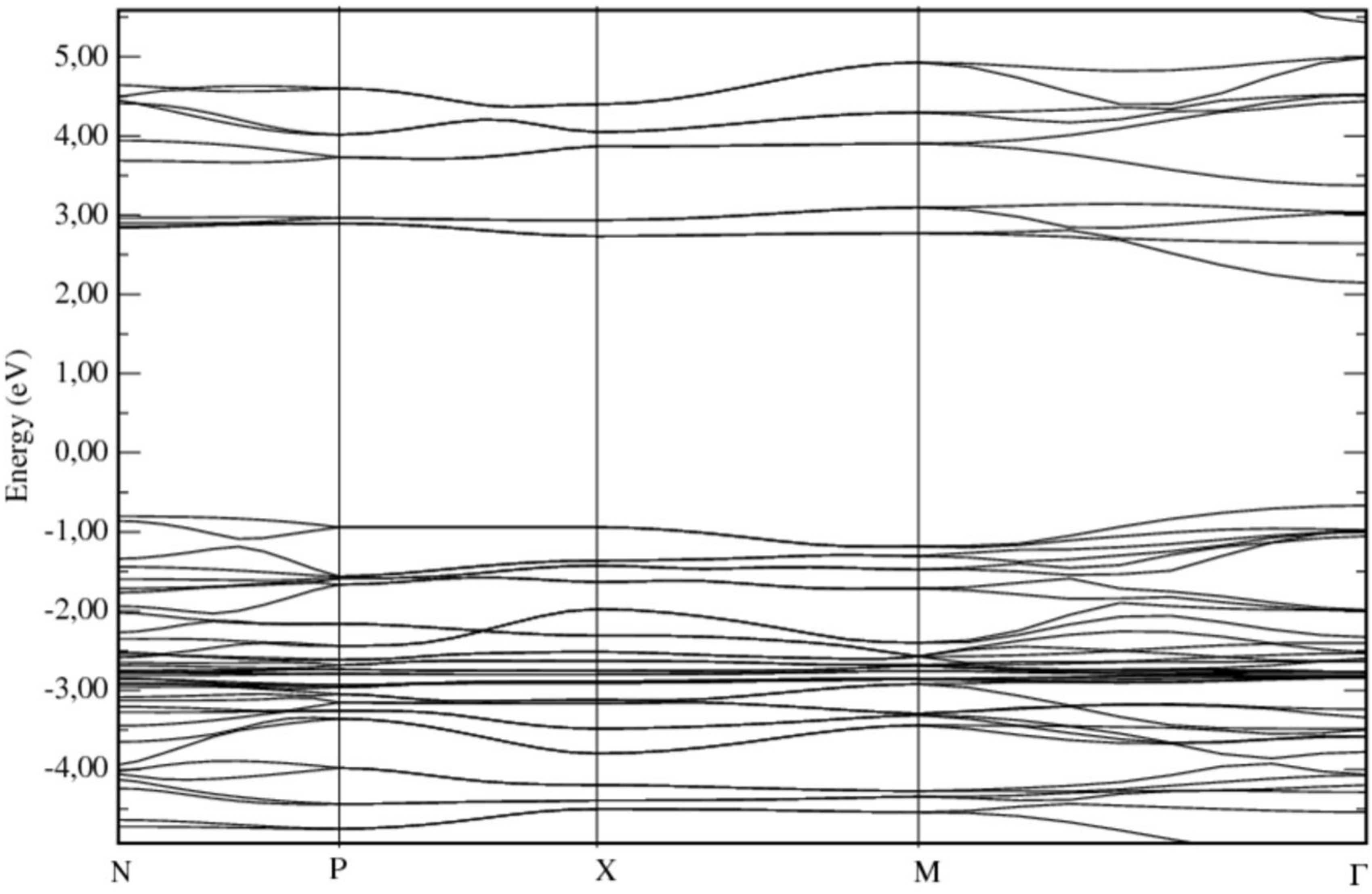

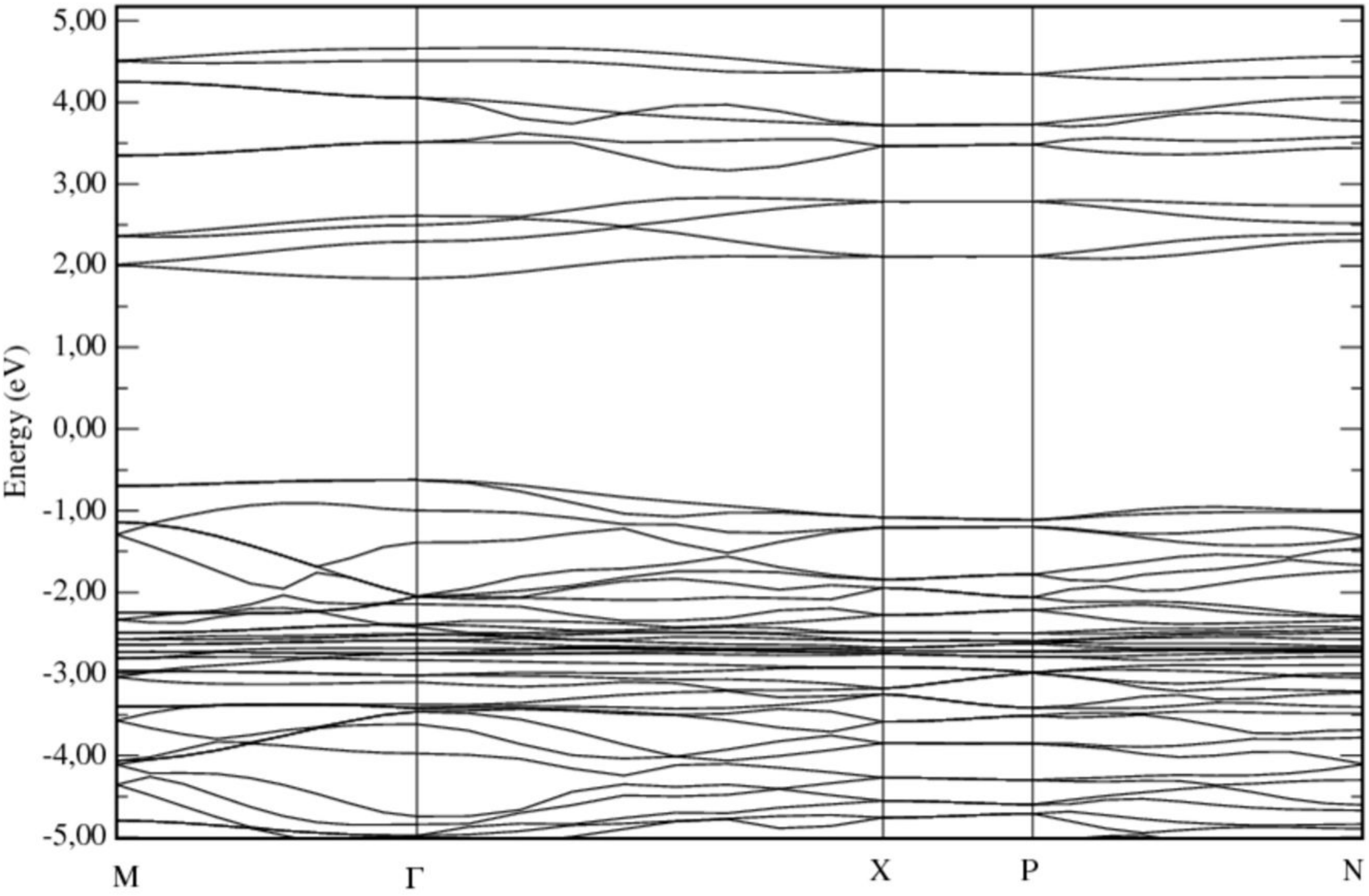

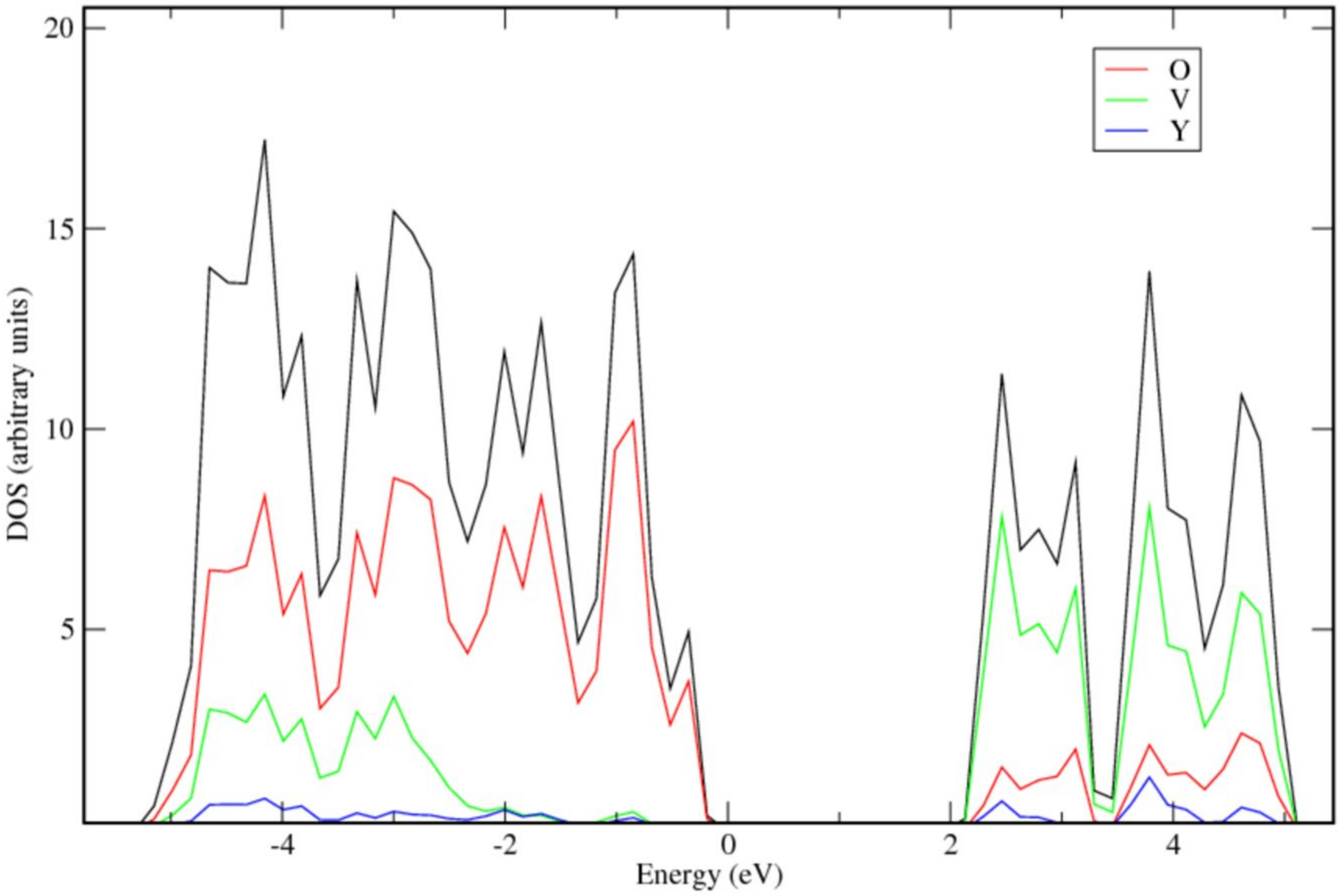

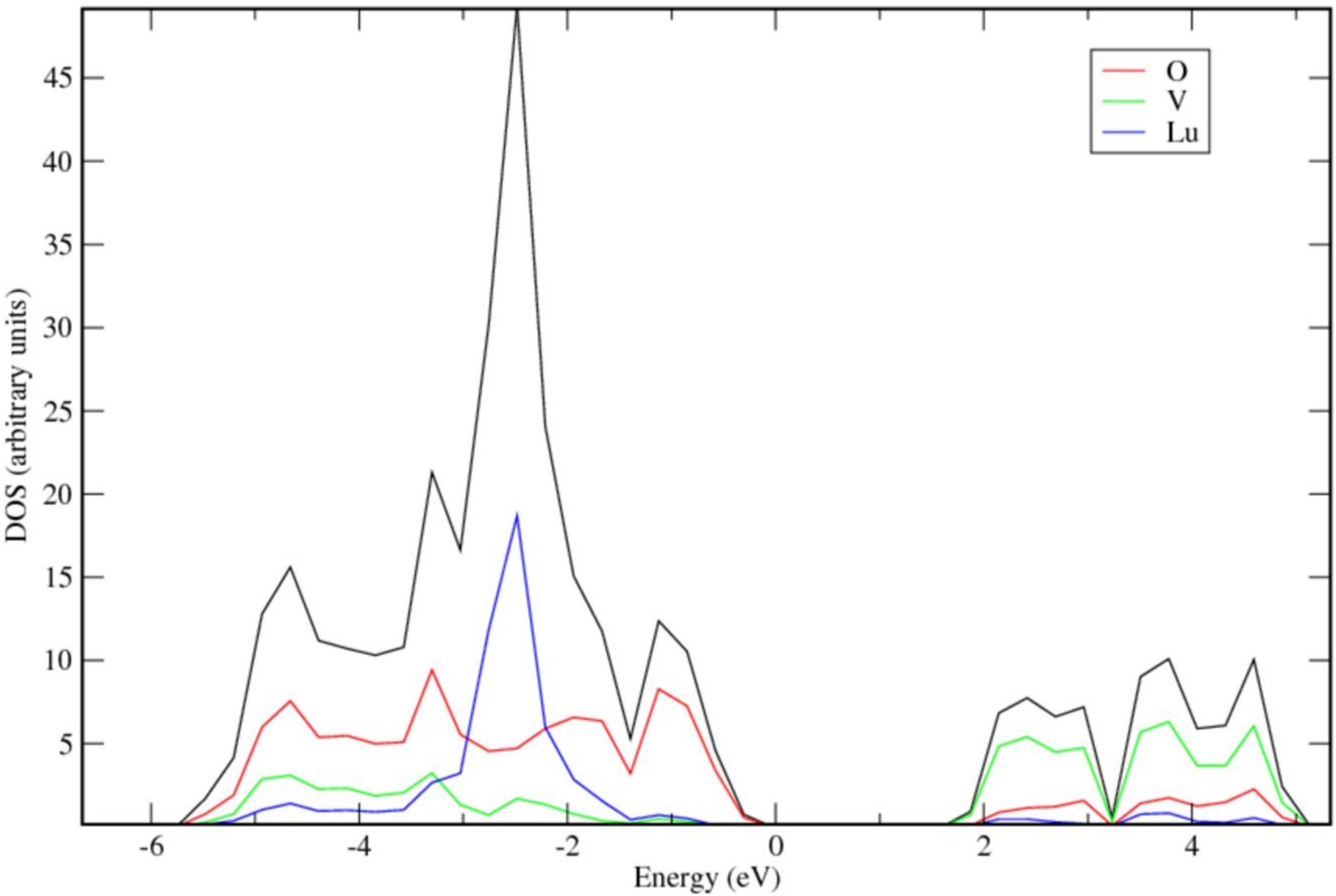

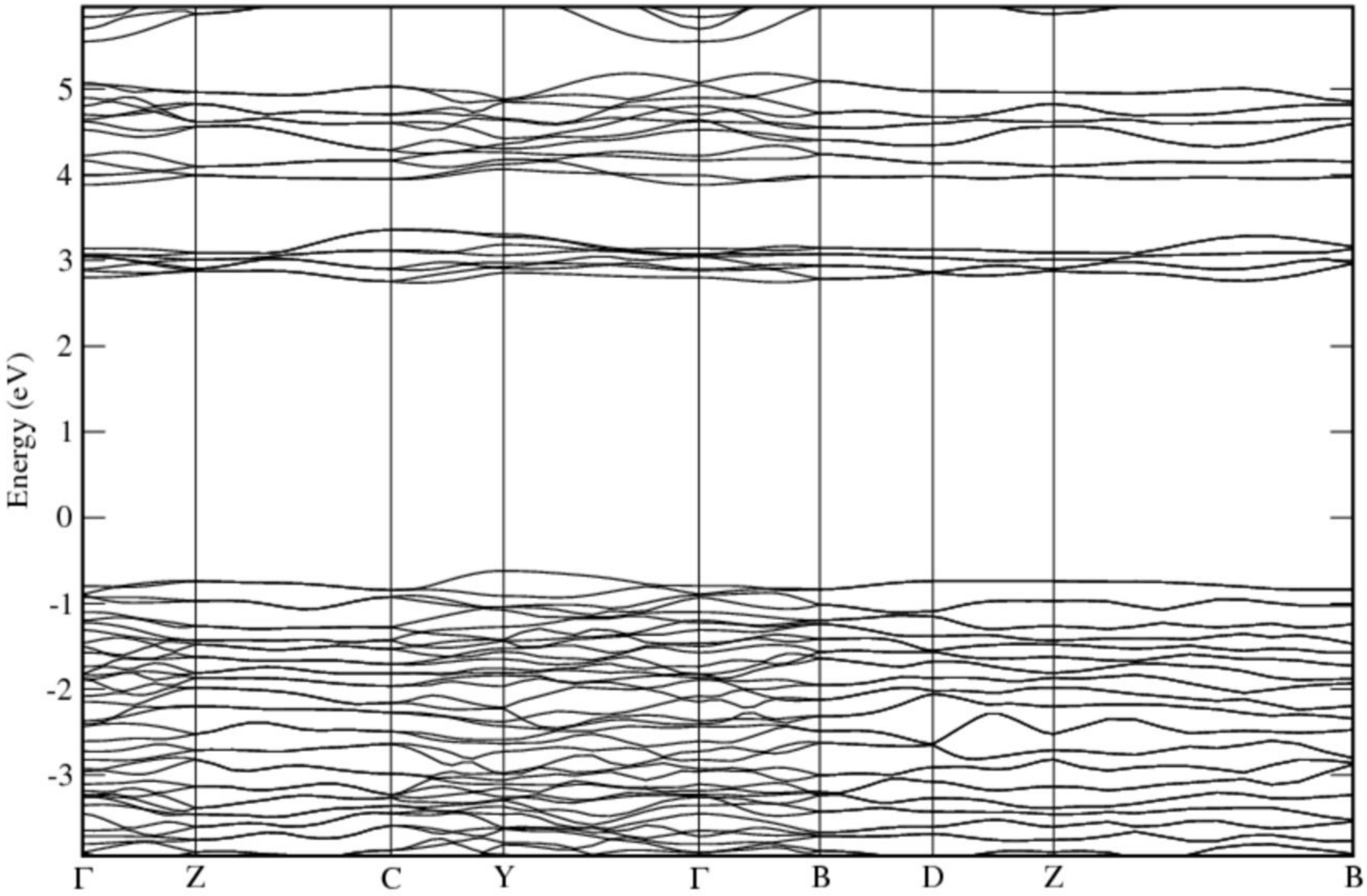

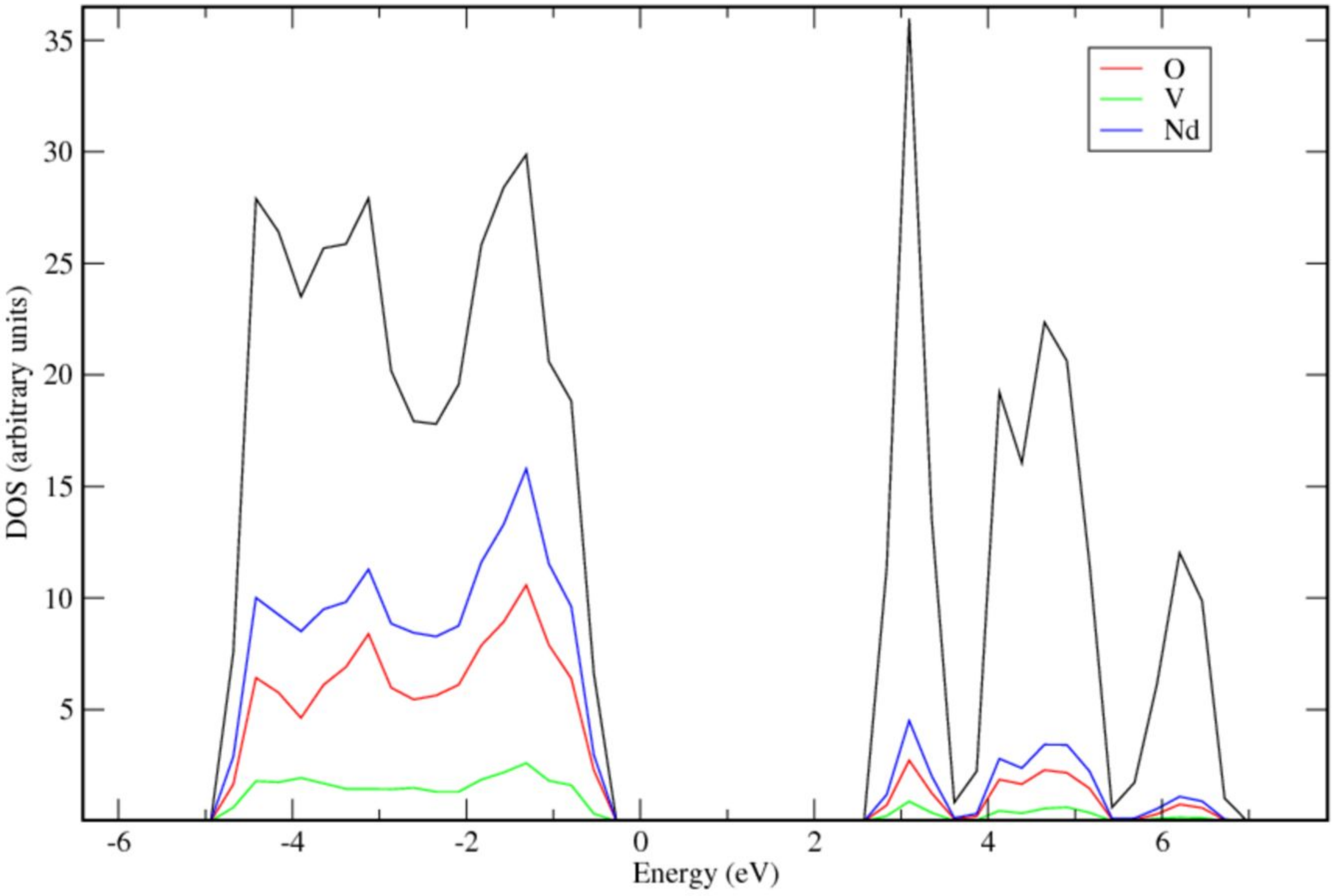